\documentclass[10pt]{iopart}

\expandafter\let\csname equation*\endcsname\relax
\expandafter\let\csname endequation*\endcsname\relax
\usepackage{amsmath} 
\usepackage[utf8]{inputenc}
\usepackage[english]{babel}
\usepackage{hyperref}
\usepackage{graphicx}
\usepackage{subcaption}
\usepackage{bm} 
\newcommand{\del}{\partial}

\begin{document}
\title[A general perturbative approach for bead-based microswimmers]{A general perturbative approach for bead-based microswimmers reveals rich self-propulsion phenomena}
\author{Sebastian Ziegler\textit{$^{a}$}, Maxime Hubert\textit{$^{a}$}, Nicolas Vandewalle\textit{$^b$} Jens Harting\textit{$^{cd}$}, and Ana-Sunčana-Smith$^{\ast}$\textit{$^{ae}$}}
\address{$^{a}$~PULS Group, Department of Physics and Interdisciplinary Center for Nanostructured Films, Friedrich-Alexander-University Erlangen-Nuremberg, Cauerstr. 3, 91058 Erlangen, Germany}
\address{$^b$~Université de Liège, GRASP, CESAM Research Unit, Institut de Physique B5a, B4000 Liège, Belgium}
\address{$^{c}$~Helmholtz Institute Erlangen-Nuremberg for Renewable Energy (IEK-11), Forschungszentrum Jülich, Fürther Str. 248, 90429 Nuremberg, Germany}
\address{$^{d}$~Department of Applied Physics, Eindhoven University of Technology, P.O. Box 513, 5600MB, Eindhoven, The Netherlands}
\address{$^{e}$~Group for Computational Life Sciences, Division of Physical Chemistry, Ruđer Bošković Institute, Bijeni\v{c}ka cesta 54, 10000 Zagreb, Croatia}
\address{$^\ast$~Email: smith@physik.fau.de}

\begin{abstract}
Studies of model microswimmers have significantly contributed to the understanding of the principles of self-propulsion we have today. However, only a small number of microswimmer types have been amenable to analytic modeling, and  further development of such approaches is necessary to identify the key features of these active systems. Here we present a general perturbative calculation scheme for swimmers composed of beads interacting by harmonic potentials, driven by an arbitrary force protocol. The hydrodynamic interactions are treated using the Oseen and Rotne-Pragner approximations. We validate our approach by using 3 bead assemblies and comparing the results with the numerically obtained full-solutions of the governing equations of motion, as well as with the existing analytic models for a linear and a triangular swimmer geometries. While recovering the relation between the force and swimming velocity, our detailed analysis and the controlled level of approximation allow us to find qualitative differences already in the far field flow of the devices. Consequently, we are able to identify a behavior of the swimmer that is richer than predicted in previous models. Given its generality, the framework can be applied to any swimmer geometry, driving protocol and beads interactions, as well as in many swimmers problems. 
\end{abstract}

\submitto{\NJP}
\maketitle

\section{Introduction}
The locomotion of swimmers at small scales has been an active area of research in recent years \cite{LaugaPowers2009}, with a variety of microswimmer models being proposed, both experimental \cite{DreyfusBaudryRoper2005, LeoniKotarBassetti2009, AhmedLuNourhani2015, GrosjeanHubertLagubeau2016, ZhengDaiWang2017, HamiltonPetrovWinlove2017, BryanShelleyParish2017, GrosjeanHubertVandewalle2018, GrosjeanHubertCollard2018, GrosjeanLagubeauDarras2015} and theoretical \cite{NajafiGolestanian2004, GolestanianAjdari2008, EarlPooleyRyder2007, BeckerKoehlerStone2003, LedesmaLoewenYeomans2012, AvronKennethOaknin2005, FriedrichJuelicher2012, BennettGolestanian2013, PandeSmith2015, RizviFarutinMisbah2018}. A number of these models aims at understanding the propulsion mechanisms of small organisms such as bacteria or algae cells, or at designing artificial microswimmers. Due to the time-independence of the Stokes equation, modelling microswimmers has turned out to be a tradeoff between as little degrees of freedom as possible and enough degrees to break the time-reversal symmetry \cite{Purcell1977}. 

A milestone in the field has been the minimalistic swimmer consisting of three spherical beads arranged in a linear fashion, introduced by Najafi and Golestanian \cite{NajafiGolestanian2004}. In their model, each neighbouring pair of beads is connected by an extendible arm of a length that is prescribed as a function of time. Calculating the swimming velocity to leading order in the extension of the arms gives rise to a simple intuition for the swimmer's speed: The displacement of the swimmer corresponding to one swimming stroke is proportional to the area enclosed by the swimmer's trajectory in the conformation space \cite{GolestanianAjdari2008}. 
This model has been used to investigate the hydrodynamic properties of microswimmers, including the flow fields they produce and their mutual interaction \cite{AlexanderPooleyYeomans2009, PooleyAlexanderYeomans2007, AlexanderPooleyYeomans2008}, as well as the interaction of a swimmer with a wall \cite{ZargarNajafiMiri2009, DaddiLisickiHoell2018}.

Despite its immense usefulness, this model suffers from the constriction of all internal degrees of freedom by the swimming stroke, namely the internal dynamical behavior of the swimmer cannot react to its surrounding. This was overcome by replacing the arms with springs and prescribing the forces acting on the beads rather than the stroke itself \cite{Felderhof2006}. In this so-called bead-spring swimmer model, the existence of a viscosity maximising the swimming velocity \cite{PandeMerchantKrueger2017a} and synchronization effects of the stroke have been discovered \cite{PandeSmith2015}. Recently, an altered version of the bead-spring model has been proposed, where the swimmer was driven by periodic changes in the equilibrium lengths of the springs \cite{YasudaHosakaKuroda2017, KurodaYasudaKomura2019}. 

The boundedness of the linear swimmer to one dimension is broken in a triangular swimmer geometry, allowing for translational as well as rotational motion \cite{DreyfusBaudryStone2005, EarlPooleyRyder2007, LedesmaLoewenYeomans2012}. This geometry has also been used to model \textit{Chlamydomonas reinhardtii} and investigate in particular the synchronization between the beating of its two flagella \cite{FriedrichJuelicher2012, PolotzekFriedrich2013, BennettGolestanian2013}. Experimentally, a triangular swimmer with intrinsic elasticity has been realized by placing ferromagnetic beads subject to an oscillating magnetic field at a water air interface \cite{GrosjeanHubertCollard2018, GrosjeanHubertVandewalle2018} and a similar system has been investigated by means of lattice Boltzmann simulations \cite{SukhovZieglerXie2019}. The full controllability of a triangular swimmer in the 2D space has been shown recently analytically \cite{RizviFarutinMisbah2018}, expanding on the use of the bead spring model \cite{PandeSmith2015}. However, the perturbative calculations in \cite{PandeSmith2015, PandeMerchantKrueger2017a, RizviFarutinMisbah2018} hold only in the limit of very large bead separations where the swimming velocity becomes extremely small. Especially since this limit is inaccessible in experiments (see  \cite{GrosjeanHubertLagubeau2016, GrosjeanHubertVandewalle2018}) as external disturbances become exuberant, an investigation of swimmers with bead separation of a few radii is required and still lacking. 

In this paper, we fill this gap by presenting a general perturbative framework to calculate the full behavior of arbitrarily shaped bead-spring swimmers, i.e. not only their stroke and swimming velocity, but also the average deformation and the produced flow field with tunable precision of the results in terms of the bead separation. Our calculation is essentially different from previous calculations as we split the equation of motion by orders of the driving force and systematically solve the long-term limit of each order \cite{Felderhof2006, PandeMerchantKrueger2017a, RizviFarutinMisbah2018}. 
Doing so significantly increases the precision of the result as the assumption of very large bead separation is not made in our calculation. The so-obtained correction is found to be small for the symmetric linear swimmer, but grows strongly when the swimmer becomes asymmetric. 
We investigate the dynamics of the triangular swimmer, and find a transient phase of rotation towards a stable steady state in which the swimmer propagates without rotating. In the state of purely translational motion, the swimmer produces a dipolar flow field, which has not been reported for the triangular bead-spring swimmer yet. 

The remainder of the article is structured as follows: in the next section we introduce the general model of bead-spring microswimmers. We proceed by analyzing the equation of motion by means of perturbation theory and present the calculation of the swimmer's velocity, flow field and the beads's trajectories step by step in section 3. Subsequently, in section 4, we apply this framework to the linear swimmer as a benchmark. Finally, in section 5, we use it to investigate the triangular swimmer in both external and internal driving. Section 6 contains the discussion and conclusion.

\section{Model}
We consider a microswimmer composed of $n$ spherical beads of radius $a_i$ in the $d$-dimensional space. Some pairs of beads are connected by linear springs which we assume here to be harmonic corresponding to the interaction potential
\begin{equation}
\phi_{ij} (\vec{R}_i, \vec{R}_j) \equiv \phi_{ij} (|\vec{R}_i - \vec{R}_j|) = \frac{k_{ij}}{2} \left(|\vec{R}_i - \vec{R}_j| - L_{ij} \right)^2. 
\label{eq:springPotential}
\end{equation}
Here, $\vec{R}_i$ and $\vec{R}_j$ are the positions of the beads, $k_{ij}$ the stiffness and $L_{ij}$ the length of the spring connecting bead $i$ and $j$ in the swimmer's mechanical equilibrium.
Our approach can also easily be adopted to more complex interaction potentials, e.g. magneto-capillary potentials (see \cite{LagubeauGrosjeanDarras2016}). Note that the springs are not tied to a certain direction but can freely rotate as the beads move in the fluid. 
The total spring potential of the device is given by 
\begin{equation}
\phi (\bm{R}) = \sum_{\mathrm{conn. \ pairs \ ij}} \phi_{ij} (\vec{R}_i, \vec{R}_j),
\end{equation}
where we sum over all pairs of beads that are connected and define $\bm{R} = (\vec{R}_1, ..., \vec{R}_n)$ as a vector with $n \cdot d$ components. 

We assume that the Reynolds number of the beads in the fluid is small and that the relaxation of the fluid takes place sufficiently fast \cite{Dhont1996} such that the fluid dynamics can be described by the Stokes equation
\begin{equation}
\nabla p(\vec{r}, t) - \eta \nabla^2 \vec{u}(\vec{r}, t) = \vec{F} (\vec{r}, t).
\end{equation}
Here, $p(\vec{r}, t)$ denotes the pressure in the fluid, $\vec{u}(\vec{r}, t)$ the velocity of the fluid, $\vec{F} (\vec{r}, t)$ the force density applied to the fluid and $\eta$ its dynamic viscosity. $\vec{r}$ and $t$ denote position vector and time, respectively. The fluid is assumed to be incompressible: 
\begin{equation}
\nabla \vec{u} (\vec{r}, t) = 0. 
\end{equation}
A point force $\vec{F}(\vec{r}) = \vec{F}_0 \delta(\vec{r} - \vec{r}_0)$, acting on the fluid at position $\vec{r}_0$, induces a fluid flow at position $\vec{r}$ given by the Oseen tensor  $\hat{T}^{\mathrm{O}} (\vec{r} -\vec{r}_0)$ as
\begin{equation}
\begin{split}
&\vec{u} (\vec{r}) = \int d^3 \vec{s} \, \hat{T}^{\mathrm{O}}(\vec{r} - \vec{s}) \vec{F}(\vec{s}) = \hat{T}^{\mathrm{O}}(\vec{r} - \vec{r}_0) \vec{F}_0 = \\ &\frac{1}{8 \pi \eta |\vec{r} - \vec{r}_0|} \left( \hat{1} + \frac{(\vec{r} - \vec{r}_0) \otimes (\vec{r} - \vec{r}_0)}{(\vec{r} - \vec{r}_0)^2} \right) \vec{F}_0,
\end{split}
\end{equation}
with $\otimes$ the tensor product and $\hat{1}$ the $d$-dimensional unity matrix. 
For a large separation between two beads in terms of their radii, the Oseen tensor is a sufficient description for their interaction. Being interested in swimmers with small bead separations, we make use of the Rotne-Prager approximation \cite{PicklPandeKoestler2017}, which is given by
\begin{equation}
\begin{split}
 \hat{T}^{\mathrm{RP}} (\vec{r} - \vec{r}_0) =  \frac{1}{8 \pi \eta |\vec{r} - \vec{r}_0|} \left( \hat{1} + \frac{(\vec{r} - \vec{r}_0) \otimes (\vec{r} - \vec{r}_0)}{(\vec{r} - \vec{r}_0)^2} \right) + \\ \frac{a_i^2 + a_j^2}{24 \pi \eta |\vec{r} - \vec{r}_0|^3} \left( \hat{1} -3  \frac{(\vec{r} - \vec{r}_0) \otimes (\vec{r} - \vec{r}_0)}{(\vec{r} - \vec{r}_0)^2} \right).
\end{split}
\label{eq:RotnePragerTensor}
\end{equation}
The mobility matrix of an ensemble of $n$ spheres is a $(n \cdot d) \times (n \cdot d)$ matrix defined in terms of $d \times d$ blocks as 
\begin{equation}
\begin{split}
\underline{\mu} (\bm{R}) \equiv \underline{\mu} ((\vec{R}_1, ..., \vec{R}_n)) = 
\begin{cases}
\begin{array}{lr}
       \hat{1}/(6 \pi \eta a_i), & \text{for } i = j\\
       \hat{T}(\vec{R}_i - \vec{R}_j), & \text{for } i \neq j
        \end{array}
\end{cases},
\end{split}
\label{eq:mobilityMatrix}
\end{equation}
with $\hat{T}$ either the Oseen or the Rotne-Prager tensor and $i,j = 1, ..., n$. In~\eqref{eq:mobilityMatrix}, we account for both, the Stokes drag as well as the interaction of the beads due to the fluid. 
The springs are assumed to be not interacting with the fluid. 

An oscillating force $\vec{E}_i (t, \bm{R})$ with fixed frequency $\omega$ is acting on each bead $i$ as
\begin{equation}
\vec{E}_i (t, \bm{R}) = A_i \vec{E}_{i} (\bm{R}) \sin(\omega t + \alpha_i),
\label{eq:driving}
\end{equation}
with $A_i$ encoding the amplitude of the driving for each bead $i$ and $\alpha_i$ the phase shift associated to each bead. The vector $\vec{E}_{i} (\bm{R})$ is dimensionless and subject to a suitable normalization that will be chosen specifically for each geometry and driving protocol. 
As indicated above, we allow for a dependence of the driving forces on the current  configuration of the swimmer. This is required if e.g. the driving forces result from spatially dependent magnetic or electric fields, or if certain demands to the driving protocol shall be fulfilled, like a constantly zero total torque for an effectively rotating swimmer.

The temporal evolution of the system is governed by the constitutive equation of the mobility matrix on the $(n \cdot d)$-dimensional configuration space of the beads as
\begin{equation}
\frac{d}{dt} \bm{R} = \underline{\mu} (\bm{R}) [\bm{E} (t, \bm{R}) + \bm{G} (\bm{R})],
\end{equation}
with 
\begin{equation}
\bm{E} (t, \bm{R}) = (\vec{E}_1 (t, \bm{R}), ..., \vec{E}_n (t, \bm{R})), \ \ 
\bm{G} (\bm{R}) = -\nabla_{\bm{R}} \phi(\bm{R}).
\end{equation}
$\nabla_{\bm{R}}$ denotes the gradient of a function with respect to all $n \cdot d$ components of $\bm{R}$. 
In view of rescaling of the equations of motion, we introduce for each family of parameters $a_i, k_{ij}, L_{ij}$ and $A_i$ a characteristic value ($a, k, L$ and $A$, respectively) and define dimensionless parameters by
\begin{equation}
a_i' := a_i/a, \ k'_{ij} := k_{ij}/k, \ L'_{ij} := L_{ij}/L, A_i' := A_i/A, 
\label{eq:aklParameter}
\end{equation}
which become 1 in the case of equal parameters of one type.

\section{Analysis}
To calculate how the swimmer behaves around a stable mechanical equilibrium, we develop a perturbative approach that allows to split the equation of motion by orders in the driving force. 
We firstly rescale the equation of motion using the characteristic length $a$ and the characteristic time $t_V = (6 \pi \eta a)/k$ \cite{PandeMerchantKrueger2017a}, denoted as viscous time. We find the effective parameters to be $\epsilon = A/(k a)$ for the rescaled driving force amplitude, $\nu = a/L$ the aspect ratio between bead radius and separation, and $\Gamma = \omega t_V$ the rescaled driving frequency. Rescaled variables are marked with an additional dash and the rescaled time is $\tau := t/t_V$. The equation of motion can then be re-expressed as
\begin{equation}
\frac{d}{d \tau} \bm{R}' = \underline{\mu}' (\bm{R}') [\epsilon \bm{E}' (\bm{R}', \tau) + \bm{G}' (\bm{R}')],
\label{eq:eomRescaledCompact}
\end{equation}
with
\begin{equation}
\bm{R}' := \frac{1}{a} (\vec{R}_1, ..., \vec{R}_n); \ \bm{E}' (\tau, \bm{R}') := \frac{1}{A} (\vec{E}_1 (\tau t_V, a \bm{R}'), ..., \vec{E}_n (\tau t_V, a \bm{R}')),
\label{eq:rescaledRAndE}
\end{equation}
\begin{equation}
\begin{split}
\underline{\mu}' (\bm{R}') = 
6 \pi \eta a \, \underline{\mu} (a \bm{R}'),
\end{split}
\end{equation}
and
\begin{equation}
\bm{G}'(\bm{R}') = \frac{1}{k a} \bm{G}(a \bm{R}'). 
\end{equation}
 
The equation of motion~\eqref{eq:eomRescaledCompact} can be solved with a perturbative approach in the vicinity of $\epsilon = 0$. Therefore, we employ a suitable power series ansatz in $\epsilon$ for the displacement out of the equilibrium
\begin{equation}
\bm{\xi}' := \bm{R}' - \bm{R}'^{\mathrm{eq}} = \epsilon \bm{\xi}'^{(1)} + \epsilon^2 \bm{\xi}'^{(2)} + \epsilon^3 \bm{\xi}'^{(3)} + ... \ ,
\label{eq:expansionDisplacement}
\end{equation}
where $\bm{R}'^{\mathrm{eq}}$ is the rescaled equilibrium configuration of the swimmer. 
A Taylor expansion of all $\bm{R}'$-dependent parts of~\eqref{eq:eomRescaledCompact} around $\bm{R}'^{\mathrm{eq}}$ yields
\begin{equation}
\begin{split}
&\frac{d}{d \tau} (\bm{R}'^{\mathrm{eq}} + \bm{\xi}') = \\
&\left[ \underline{\mu}' (\bm{R}'^{\mathrm{eq}}) + \del_\alpha \underline{\mu}' (\bm{R}'^{\mathrm{eq}}) \bm{\xi}'^\alpha + \frac{1}{2} \del_\alpha \del_\beta \underline{\mu}' (\bm{R}'^{\mathrm{eq}}) \bm{\xi}'^\alpha \bm{\xi}'^\beta + ... \right]\times \\
&\left[\left( \epsilon \bm{E}' (\bm{R}'^{\mathrm{eq}}, \tau) + \epsilon \del_\alpha \bm{E}' (\bm{R}'^{\mathrm{eq}}, \tau) \bm{\xi}'^\alpha + \frac{\epsilon}{2} \del_\alpha \del_\beta \bm{E}' (\bm{R}'^{\mathrm{eq}}, \tau) \bm{\xi}'^\alpha \bm{\xi}'^\beta + ... \right) \  \right. \\
&\left.  + \left( \bm{G}' (\bm{R}'^{\mathrm{eq}}) + \del_\alpha \bm{G}' (\bm{R}'^{\mathrm{eq}}) \bm{\xi}'^\alpha + \frac{1}{2} \del_\alpha \del_\beta \bm{G}' (\bm{R}'^{\mathrm{eq}}) \bm{\xi}'^\alpha \bm{\xi}'^\beta + ... \right) \right].
\label{eq:eomRescaledExpanded}
\end{split}
\end{equation}
$\alpha, \beta$ are summed over when appearing repeatedly and go from 1 to $n \cdot d$, $\bm{\xi}'^\alpha$ denotes the $\alpha$-th component of $\bm{\xi}'$, and $\del_\alpha$ the derivative with respect to the $\alpha$-th component of $\bm{R}'$. 
The $\tau$-derivative of $\bm{R}'^{\mathrm{eq}}$ is zero as well as the spring forces evaluated in the equilibrium $\bm{G}' (\bm{R}'^{\mathrm{eq}})$. 

We proceed by replacing $\bm{\xi}'$ in \eqref{eq:eomRescaledExpanded} 
by its power series in $\epsilon$ \eqref{eq:expansionDisplacement}. Ordering and splitting the resulting equation by powers of $\epsilon$ yields a vectorial equation for each order $p = 1, 2, ...$ in $\epsilon$. One finds that each of them is of the generic form
\begin{equation}
\frac{d}{d \tau} \bm{\xi}'^{(p)} = \underline{K}' \bm{\xi}'^{(p)} + \bm{S}'^{(p)} (\tau),
\label{eq:eomRescaledSplit}
\end{equation}
with 
\begin{equation}
\underline{K}' = \underline{\mu}' (\bm{R}'^{\mathrm{eq}}) \circ (\nabla_{\bm{R}'} \bm{G}') (\bm{R}'^{\mathrm{eq}}),
\label{eq:definitionK}
\end{equation}
where $\circ$ denotes the matrix multiplication and $(\nabla_{\bm{R}'} \bm{G}') (\bm{R}'^{\mathrm{eq}})$ is the Jacobian matrix of the spring forces, evaluated at the swimmer's equilibrium. $\bm{S}'^{(p)}$ is a term that only depends on the displacements $\bm{\xi}'^{(1)}, ..., \bm{\xi}'^{(p-1)}$ and on the derivatives (of first and higher order) of $\underline{\mu}', \bm{E}'$ and $\bm{G}'$, evaluated at $\bm{R}'^{\mathrm{eq}}$. Since $\bm{S}'^{(p)}$ does not depend on $\bm{\xi}'^{(p)}$, it can be considered as a source term that is known assuming \eqref{eq:eomRescaledSplit} are solved in ascending order in $p$. Consequently, the first two source terms read
\begin{equation}
\bm{S}'^{(1)} (\tau) = \underline{\mu}' (\bm{R}'^{\mathrm{eq}}) \, \bm{E}' (\bm{R}'^{\mathrm{eq}}, \tau), 
\label{eq:sourceTermO1}
\end{equation}
\begin{multline}
\bm{S}'^{(2)} = \\ 
\underline{\mu}'  (\bm{R}'^{\mathrm{eq}})  \left( \frac{1}{2} \del_\alpha \del_\beta \bm{G}' (\bm{R}'^{\mathrm{eq}}) \bm{\xi}'^{(1) \alpha}  \bm{\xi}'^{(1) \beta} + \del_\alpha \bm{E}' (\bm{R}'^{\mathrm{eq}}, \tau) \bm{\xi}'^{(1) \alpha} \right) + \\ 
\del_\alpha \underline{\mu}' (\bm{R}'^{\mathrm{eq}}) \bm{\xi}'^{(1) \alpha} \left(\bm{E}' (\bm{R}'^{\mathrm{eq}}, \tau) + \del_\beta \bm{G}' (\bm{R}'^{\mathrm{eq}}) \bm{\xi}'^{(1) \beta} \right).
\label{eq:sourceTermO2}
\end{multline}

The matrix $\underline{K}'$ maps the displacement vector $\bm{\xi}'$ to the velocity vector $(d/dt) \bm{R}' = (d/dt) \bm{\xi}' = \underline{K}' \bm{\xi}'$ that emerges from the displacement 
in the situation of zero external forces.
We assume the mobility matrix $\underline{\mu}' (\bm{R}'^{\mathrm{eq}})$ to be positive definite and symmetric. The gradient of the spring forces, $(\nabla_{\bm{R}'} \bm{G}') (\bm{R}'^{\mathrm{eq}})$, has to be negative semi-definite due to the stability of the swimmer's equilibrium and symmetric as it is the Hessian of the spring potential. For the latter matrix, we distinguish between eigenvalues being zero, associated to translations and rotations of the whole swimmer, and negative eigenvalues, associated to deformations of the swimmer. Then, it is easy to show that, similarly to $(\nabla_{\bm{R}'} \bm{G}') (\bm{R}'^{\mathrm{eq}})$, the matrix $\underline{K}'$, defined by~\eqref{eq:definitionK}, is diagonalizable and has only non-positive eigenvalues, where zero as an eigenvalues is associated to translations/rotations and negative eigenvalues are associated to internal degrees of freedom \cite{HornJohnson1985}.

The explicit way to solve a set of differential equations like~\eqref{eq:eomRescaledSplit}, accounting for certain initial conditions, is to split the initial conditions by orders of $\epsilon$, find the full solution for each order, and adjust it to these initial conditions by a suitable choice of the parameters in the homogeneous solution.  
Knowing that $\underline{K}'$ has only non-positive eigenvalues $\lambda_{(\alpha)}$, the solution to the homogeneous counterpart to~\eqref{eq:eomRescaledSplit} ($\bm{S}'^{(p)} = 0$) can be written as 
\begin{equation}
\bm{\xi}'_{hom} = \sum_\alpha \bm{X}_\alpha \exp(\lambda_{(\alpha)} \tau),
\label{eq:solutionHomogeneousEq}
\end{equation} 
with $\bm{X}_\alpha$ a suitably scaled eigenvector of $\underline{K}'$ corresponding to the eigenvalue $\lambda_{(\alpha)}$. 
For translational and rotational degrees of freedom, one has $\lambda_{(\alpha)} = 0$ and hence a constant solution. The displacements corresponding to the internal degrees of freedom are exponentially decaying and hence go to zero for large $\tau$. Therefore, the homogeneous solution describes the relaxation of the swimmer in the absence of driving from arbitrary initial conditions. Consequently, all solutions to the inhomogeneous equations ($\bm{S}'^{(p)} \neq 0$) differ only by the constant homogeneous solution for large $\tau$. It suffices to find a single arbitrary solution to the full~\eqref{eq:eomRescaledSplit} in order to determine the swimming velocity and the deformation of the swimmer. For a fixed source term, the swimmer's behavior is hence independent of the initial conditions for large $\tau$. 
The eigenvalues corresponding to internal degrees of freedom are of the order of 1, such that the displacements corresponding to the internal degrees of freedom decay exponentially with a characteristic time of the order of $t_V$. 
In the following calculations, we are interested in the behavior of the swimmer on times scales larger than $t_V$. Therefore, we will neglect all terms decaying exponentially at time scale $t_V$ arising from the initial conditions in the displacements $\bm{\xi}'^{(1)}, ..., \bm{\xi}'^{(p-1)}$ when calculating the source term of each order $p$. 

We point out here that if the system is not invariant under the translational or rotational degree of freedom, e.g. because the driving forces are not invariant under these transformations, the source term $\bm{S}'^{(p)} (\tau)$ explicitly depends on those degrees of freedom. Via that pathway, the initial conditions may actually have an impact on the swimmer's behavior. This will be the case for the triangular swimmer in external driving as it will be discussed in section 5.1.

The rescaled flow field $\vec{u}'_{\mathrm{fluid}} (\vec{r}')$ generated by the swimmer, expressed in dependence of the rescaled position $\vec{r}'$, is given by 
\begin{equation}
\vec{u}'_{\mathrm{fluid}} (\vec{r}') = \sum_{i = 1}^n \hat{T}'(\vec{r}' - \vec{R}'_i) \left( \epsilon \bm{E}'(\bm{R}', \tau) + \bm{G}'(\bm{R}') \right)^i, 
\label{eq:flowField}
\end{equation}
with $\hat{T}'(\vec{x}') := 6 \pi \eta a \hat{T}(a \vec{x}')$ the rescaled Oseen/Rotne-Prager tensor and $(\bullet)^i$ the $i$-th part in a decomposition of $\bullet$ into $n$ partial vectors of length $d$, i.e. the components associated to the $i$-th bead. Given the solutions of~\eqref{eq:eomRescaledSplit} up to order $p$, the flow field can be calculated up to the same order $p$ in $\epsilon$ by expanding~\eqref{eq:flowField}. We here state the explicit expression up to the second order in $\epsilon$: 
\begin{equation}
\begin{split}
\vec{u}_{\mathrm{fluid}}'(\vec{r}') = \epsilon \sum_{i = 1}^n \hat{T}'(\vec{r}' - \vec{R}'^{\mathrm{eq}}_i) \left( \bm{E}'(\bm{R}'^{\mathrm{eq}}, \tau) + \del_\alpha \bm{G}' (\bm{R}'^{\mathrm{eq}}) \bm{\xi}'^{(1) \alpha} \right)^i + \\
\epsilon^2 \left[ \sum_{i = 1}^n \bm{\xi}^{\prime (1)}_i \cdot \nabla_{\vec{R}'_i} \hat{T}' (\vec{r}' - \vec{R}'_i)\rvert_{\vec{R}'_i = \vec{R}'^{\mathrm{eq}}_i}  \left( \bm{E}'(\bm{R}'^{\mathrm{eq}}, \tau) + \del_\alpha \bm{G}' (\bm{R}'^{\mathrm{eq}}) \bm{\xi}'^{(1) \alpha} \right)^i + \right. \\
\sum_{i = 1}^n \hat{T}'(\vec{r}' - \vec{R}'^{\mathrm{eq}}_i) \left( \del_\alpha \bm{E}' (\bm{R}'^{\mathrm{eq}}, \tau) \bm{\xi}'^{(1) \alpha} + \del_\alpha \bm{G}' (\bm{R}'^{\mathrm{eq}}) \bm{\xi}^{\prime (2) \alpha} + \right. \\
\left. \left. \frac{1}{2} \del_\alpha \del_\beta \bm{G}' (\bm{R}'^{\mathrm{eq}}) \bm{\xi}'^{(1) \alpha} \bm{\xi}'^{(1) \beta} \right)^i \right] + \epsilon^3... ,
\end{split}
\label{eq:flowFieldPerturbative}
\end{equation}
with $\nabla_{\vec{R}'_i}$ the gradient with respect to $\vec{R}'_i$. 
Note that having solved for the displacements in advance, we simply need to insert them into~\eqref{eq:flowFieldPerturbative} to obtain the flow fields produced by the swimmer. 

\section{Linear three-bead swimmer}
\begin{figure}
\centering
\includegraphics{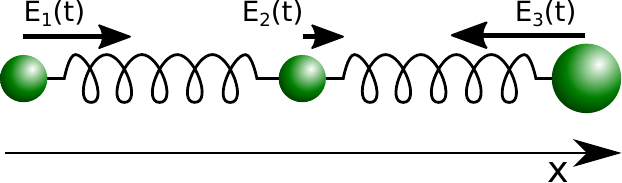}
\caption{Sketch of the linear three-bead swimmer with driving forces $E_i(t)$.}
\label{fig:sketchLinearSwimmer}
\end{figure}

The simplest one-dimensional bead-spring swimmer able to swim at low Reynolds number is the linear three-bead swimmer (figure~\ref{fig:sketchLinearSwimmer}). A swimmer with two beads comes with only one internal degree of freedom which is not sufficient to break the time reversal symmetry \cite{Purcell1977}. 
It has been studied in detail in previous analytical works \cite{Felderhof2006, PandeSmith2015, PandeMerchantKrueger2017a}, where two similar perturbative calculations were performed. In the latter works \cite{PandeSmith2015, PandeMerchantKrueger2017a}, the results for the swimming velocity were expanded and truncated in orders of $\nu$. With the precision of the Oseen tensor, this calculation does not allow for a predictive result at higher order than the leading order $\nu^2$. 

We will show that the result of our approach for the swimming velocity employing the Oseen tensor is correct up to order $\nu^3$ and that using the Rotne-Prager tensor hydrodynamics has an impact at orders $\nu^4$ and higher. Furthermore, we find that our approach coincides at leading order $\nu^2$ to the aforementioned previous results but differs at order $\nu^3$, explaining why our results also hold at order $\nu^3$. 

The swimmer consists of three beads for which we choose radii $a_1 = a_2 = a$ and $a_3$, respectively \cite{Felderhof2006}. The beads are connected by two identical harmonic springs of stiffness $k$ and equilibrium length $L$. The driving forces $\vec{E}_i (t)$, specified as 
\begin{equation}
\vec{E}_1 (t) = A \sin(\omega t), \ \vec{E}_2 (t) =  A \sin(\omega t + \alpha), \ \vec{E}_3(t) = -\vec{E}_1 (t) -  \vec{E}_2 (t),
\end{equation}
act on the beads, ensuring that the total force vanishes. 
The second derivative of $\bm{G}'(\bm{R}')$ vanishes, because the swimmer is restricted to one dimension. Furthermore, the driving forces are not spatially dependent and hence all their spatial derivatives vanish too. 

We here calculate the displacement $\bm{\xi}'$ up to second order in $\epsilon$, pointing out that the displacement to higher order can be obtained analogously. The source term $\bm{S}'^{(1)} (\tau)$, \eqref{eq:sourceTermO1}, composes of purely oscillating contributions with frequency $\Gamma$,
\begin{equation}
\bm{S}'^{(1)} (\tau) = \bm{S}'^{(1)}_{s1} \sin(\Gamma \tau) + \bm{S}'^{(1)}_{c1} \cos(\Gamma \tau), 
\end{equation}
with the indices $s1$ and $c1$ denoting first the correspondence to $\sin$ or $\cos$ and second indicating the argument of the trigonometric function in multiples of $\Gamma \tau$. 
To safe efforts later, we calculate here the solution for a more general source term given by 
\begin{equation}
\bm{S}' (\tau) = \bm{S}'_{sf} \sin(f \Gamma \tau) + \bm{S}'_{cf} \cos(f \Gamma \tau),
\label{eq:assumptionSourceTerm}
\end{equation}
with $f$ an arbitrary positive integer. 
Given the linear nature of~\eqref{eq:eomRescaledSplit}, a suitable ansatz for the displacement is $\bm{\xi}' (\tau) = \bm{\xi}'_{sf} \sin(f \Gamma \tau) + \bm{\xi}'_{cf} \cos(f \Gamma \tau)$. The resulting solutions to~\eqref{eq:eomRescaledSplit} read
\begin{equation}
\begin{split}
{\bm{\xi}'}_{sf} =  \left(f^2 \Gamma^2 \hat{\underline{1}} + \underline{K}'^2 \right)^{-1} (f \Gamma \bm{S}'_{cf} - \underline{K}' \bm{S}'_{sf}), \\
{\bm{\xi}'}_{cf} = - \left(f^2 \Gamma^2 \hat{\underline{1}} + \underline{K}'^2 \right)^{-1}  (f \Gamma \bm{S}'_{sf} + \underline{K}' \bm{S}'_{cf}).
\label{eq:solOscillatingContributions}
\end{split}
\end{equation}
These results are in full agreement with \cite{Felderhof2006}. 

\begin{figure}[h]
\centering
\includegraphics[scale=1]{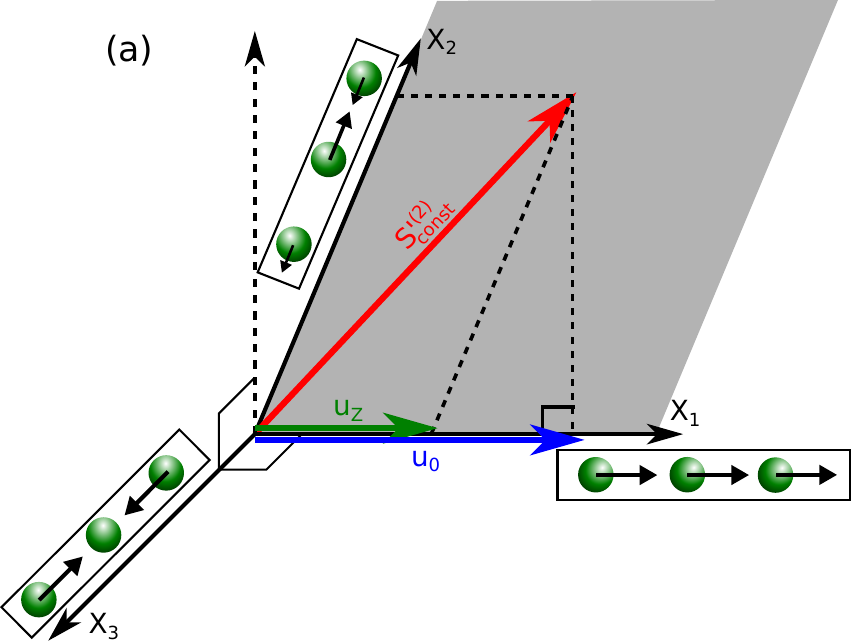}
\includegraphics[scale=1.]{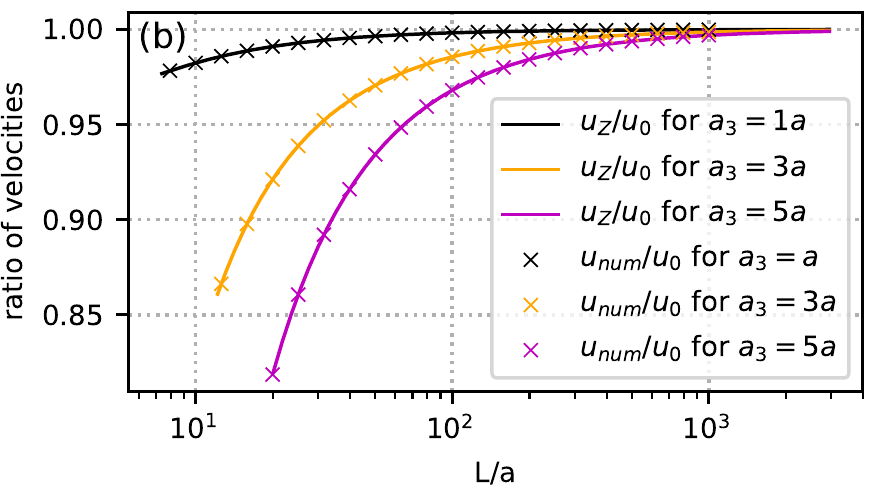}
\caption{Comparison between the perturbative calculation from \cite{Felderhof2006} and our approach, both to the second order in $\epsilon$. 
(a) Displacements of the beads, as indicated by blunt arrows centred at the beads, are decomposed into a translational mode ($X_1$) and two internal modes $X_2$ and $X_3$. The mode $X_1$ is associated to the swimming velocity. In $X_2$ mode the central bead is in counterphase relative to the outer beads, two of which are oscillating in phase. The $X_3$ mode is associated with the counterphase motion of the two outer beads, while the central bead is at rest. These modes are orthogonal only for $\nu=0$. The graph shows the decomposition of the constant contribution to the second order source term responsible for setting the swimming velocity \eqref{eq:swimmingVelFromSource}. In \cite{Felderhof2006}, the translation is a result of the orthogonal projection on the $X_1$ axis (giving rise to $u_0$), while the current approach relies on the proper decomposition, and provides $u_Z$.
(b) Comparison of our analytical and numerical results to previous outcomes. Analytical and numerical results are shown as fractions of the result from \cite{Felderhof2006}, showing increasing difference for increasing asymmetry of the swimmer. Parameters: $\epsilon=1/10, \Gamma = 1, \alpha = \pi/2$.}
\label{fig:decompositionConstSource}
\end{figure}

Having calculated the first order displacement in $\epsilon$, we proceed by calculating the second order source term~\eqref{eq:sourceTermO2}. We find that it contains oscillating terms of the frequency $2 \Gamma$ and a constant contribution:
\begin{equation}
\bm{S}'^{(2)} (\tau) = \bm{S}'^{(2)}_{s2} \sin(2 \Gamma \tau) + \bm{S}'^{(2)}_{c2} \cos(2 \Gamma \tau) + \bm{S}'^{(2)}_{const}.  
\label{eq:sourceTermOrder2}
\end{equation}
Again, the linearity of~\eqref{eq:eomRescaledSplit} allows us to calculate its solution for each summand in~\eqref{eq:sourceTermOrder2} separately and to add up the results to obtain a full solution. 
Firstly, the oscillating parts in the source term yield oscillating contributions to the $\epsilon^2$-displacement (see~\eqref{eq:assumptionSourceTerm},\eqref{eq:solOscillatingContributions}), which contribute to the stroke of the swimmer. 
Secondly,~\eqref{eq:eomRescaledSplit} with the constant source term alone can be easily solved by expressing the source term in the eigenbasis of $\underline{K}'$, where this matrix becomes diagonal and the equations separate. We find for each eigenvalue $\lambda_{(\alpha)}$, $\alpha = 1, ..., n \cdot d$:
\begin{equation}
\frac{d}{d \tau} \overline{\bm{\xi}}'^{(2) \alpha} (\tau) = \lambda_{(\alpha)}  \overline{\bm{\xi}}'^{(2) \alpha} (\tau) + \overline{\bm{S}}'^{(2) \alpha}_{const},
\label{eq:eomSplitInEigenbasis}
\end{equation}
with the overline indicating the expression of a vector in the eigenbasis of $\underline{K}'$. 
For the translational and rotational degrees of freedom with respect to $\underline{K}'$ we have $\lambda_{(\alpha)} = 0$ and the solution is a linear function in time $\tau$ plus a constant which we neglect here, as it is determined by the choice initial conditions:
\begin{equation}
\overline{\bm{\xi}}'^{(2) \alpha} (\tau) = \overline{\bm{S}}'^{(2) \alpha}_{const} \cdot \tau.
\label{eq:swimmingVelFromSource}
\end{equation}
For the internal degrees of freedom, $\lambda_{(\alpha)} < 0$, the solution to~\eqref{eq:eomSplitInEigenbasis} is constant in time plus an exponentially decaying term that we neglect since we are only interested in the limit of $\tau \gg t_V$:
\begin{equation}
\overline{\bm{\xi}}'^{(2) \alpha} (\tau) = -\frac{1}{\lambda_{(\alpha)}} \overline{\bm{S}}'^{(2) \alpha}_{const}.
\end{equation}
In effect, this describes a deformation of the swimmer such that the beads do not oscillate around their mechanical equilibrium but around a different, deformed configuration. 

Analyzing the eigensystem of $\underline{K}'$, we find that the linear three-bead spring swimmer with equal radii and restricted to one dimension has one translational mode $X_1$ and two internal eigenmodes $X_2$ and $X_3$ (see figure~\ref{fig:decompositionConstSource} a) with 
\begin{equation}
X_1 = (1, 1, 1),\ X_2 = (1, -2 + \frac{6 \nu}{9 \nu - 4}, 1),\ X_3 = (1, 0, -1).
\end{equation} 
The mode $X_3$ is orthogonal (with respect to the standard scalar product) to $X_1$ and $X_2$. The modes $X_1$ and $X_2$ are in general not orthogonal to each other, but become orthogonal for $\nu \to 0$. In our calculation, the swimming velocity can be read off from the component parallel to $X_1$ in the decomposition of $\bm{S}^{\prime (2)}_{const}$ in terms of the eigenvectors of $\underline{K}$ (green arrow in figure~\ref{fig:decompositionConstSource} a). 

In a previous work \cite{Felderhof2006}, the swimming velocity was effectively calculated as 
\begin{equation}
u'_0 = \frac{1}{\sum_p a_p} \sum_i a_i \bm{S}'^{(2) i}_{const},
\label{eq:FelderhofProjection}
\end{equation}
with $\bm{S}'^{(2) i}_{const}$ being the part of the constant source term associated to bead $i$. 
In the case of equal radii, this calculation is equivalent to an orthogonal projection of the source term onto the vector $X_1$ and reading off the velocity from the projected source term in multiples of $X_1$ (blue arrow in figure~\ref{fig:decompositionConstSource} a). 
Due to $X_2 \perp X_1$ for $\nu \to 0$, the axis projected onto is in this limit orthogonal to the two internal modes and both the projection and the decomposition of the source term yield the same result for the swimming velocity. This explains why both calculations agree for $\nu \to 0$, but also why they differ for finite values of $\nu$. 
This can also be seen in the explicit ratio between our result $u_Z$, given by 
\begin{equation}
u^{\mathrm{O, lin}}_Z = B \frac{\sin (\alpha ) \left(16 \Gamma ^2+3 (3 \nu -4) (7 \nu -4)\right)-4 \Gamma  (9 \nu -4) (2 \cos (\alpha )+1)}{(7 \nu -4) \left(16 \Gamma^2+9 (4-7 \nu )^2\right) \left(16 \Gamma ^2+(4-3 \nu )^2\right)},
\label{eq:linearThreeSwimmingVel}
\end{equation}
with $B = (a/t_V) \, \epsilon^2 \Gamma  \nu ^2 \,  (9 \nu  (21 \nu -22)+56)$, and the result $u_0$ stated in \cite{Felderhof2006}. This ratio \eqref{eq:ratioFelderhofMe} converges to 1 for $\nu \to 0$:   
\begin{equation}
\frac{u^{\mathrm{O, lin}}_Z}{u_0} = \frac{112 - 396 \nu + 378 \nu^2}{112 - 376 \nu + 315 \nu^2}.
\label{eq:ratioFelderhofMe}
\end{equation}

The differences discussed above stem in part from the fact that the original perturbative approach \cite{Felderhof2006} calculates the oscillations of the beads around the undisturbed swimmer shape. In contrast, in the current scheme, the swimmer's average shape and the oscillations of the beads around it are obtained simultaneously. Actually, both the mean deformation and the swimming velocity (the translation mode) arise from the constant contribution to the second order source term $\bm{S}^{\prime (2)}_{const}$~\eqref{eq:sourceTermOrder2}. 
The distances $d_1$ between bead 1 and 2 and $d_2$ between bead 2 and 3 in the swimmer's mean configuration, around which the beads oscillate harmonically, are given up to order $\epsilon^2$ and for $a_3 = a$ by
\begin{equation}
d_1 = L + \frac{20 - 63 \nu}{56 + 9 \nu (-22 + 21 \nu)} \cdot t_V \cdot u_Z,
\end{equation}
and 
\begin{equation}
d_2 = L + \frac{-20 + 63 \nu}{56 + 9 \nu (-22 + 21 \nu)} \cdot t_V \cdot u_Z.
\end{equation}
Hence, the ratio between deformation and swimming velocity, both of order $\epsilon^2$, is a simple geometrical factor. Also, the amplitude of the deformation obeys a similar frequency dependence as the swimming velocity itself and decays similarly as $1/\nu^2$ for large bead separations. 

The comparison to numerical calculations, done by numerically integrating the equation of motion~\eqref{eq:eomRescaledCompact}, shows a very neat agreement between our result and the numerics with errors below 0.1 \% (figure~\ref{fig:decompositionConstSource} b). 
Comparing our result with the one obtained in \cite{Felderhof2006}, we find for $a_3 = a$ (i.e. equal radii) and $\nu \approx 1/10$ small differences in the range of percents, but the difference increases drastically for increasing values of $a_3$ (figure~\ref{fig:decompositionConstSource} b), i.e. when the swimmer becomes asymmetric. 
We observe that for $a_3 \neq a$, all pairs of eigenvectors are in general no more orthogonal, even in the limit $\nu \to 0$. Also,~\eqref{eq:FelderhofProjection} describes no more a projection onto $X_1$ but onto $(1, 1, a_3)$. This explains why the difference between $u_0$ and $u_Z$ grows for increasing $a_3$ in figure~\ref{fig:decompositionConstSource} b, yet in the limit $\nu \to 0$ both still agree independently of $a_3$. 

Despite this quantitative difference to previous results \cite{Felderhof2006}, we recover the typical dependencies for bead-spring swimmers which have been reported previously \cite{Felderhof2006, PandeSmith2015, PandeMerchantKrueger2017a, SukhovZieglerXie2019}: The swimming velocity scales with the square of the driving forces for small amplitudes, the swimming speed becomes maximal in the vicinity of $t_V^{-1}$ and decays as $1/L^2$ for large bead separations.

The Rotne-Prager approximation (see Appendix) has only an impact onto the swimming velocity at orders $\nu^4$ and higher: 
Using the Rotne-Prager tensor instead of the Oseen tensor yields an additional term to the mobility matrix scaling as $1/r^3$ (see~\eqref{eq:RotnePragerTensor}), with $r$ the distance between the beads. This results in additional terms to $\underline{\mu}'  (\bm{R}'^{\mathrm{eq}})$ and $\del_\alpha \underline{\mu}' (\bm{R}'^{\mathrm{eq}})$, scaling with $\nu^3$ and $\nu^4$, respectively. A closer investigation of the second order source term~\eqref{eq:sourceTermO2} shows that the factor multiplied to $\underline{\mu}'  (\bm{R}'^{\mathrm{eq}})$ scales linearly in $\nu$, such that the additional term due to the Rotne-Prager extension has an impact at order $\nu^4$ and higher on the source term $\bm{S}^{\prime (2)}$ and likewise on the swimming velocity.

\section{Triangular swimmer}
\subsection{External driving}
Triangular bead-spring swimmers have been studied in detail recently \cite{RizviFarutinMisbah2018, RizviFarutinMisbah2018b}, where the employed driving protocol prescribes forces on each bead parallel to the adjoining sides of the triangle. 
By varying the amplitudes of and the phase shifts between the driving forces, the swimmer can be steered on arbitrary trajectories. Both, translational and rotational motion were shown to scale with the square of the driving force. Also here, the perturbative approach used in \cite{RizviFarutinMisbah2018} does only hold in the limit $\nu \to 0$.

\begin{figure}
\centering
\includegraphics{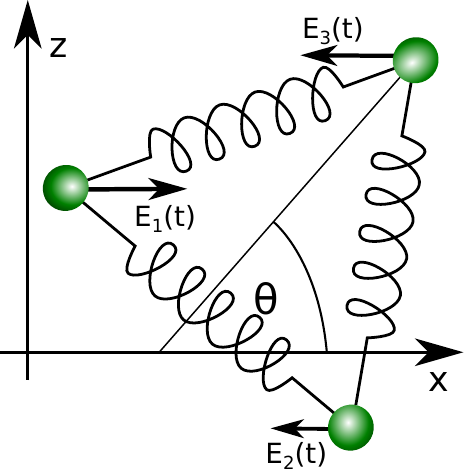}
\caption{Sketch of the triangular swimmer.}
\label{fig:sketchTriangularSwimmer}
\end{figure}
The triangular swimmer is composed of three spherical beads (radius $a$) connected by identical springs of equilibrium length $L$ and spring constant $k$. All beads are placed in the $x$-$z$-plane with orientation $\theta$ the angle between the connection of bead 3 to the middle between bead 1 and 2 and the $x$-axis (figure~\ref{fig:sketchTriangularSwimmer}). 
Numerous experimental realizations of microswimmers rely on an external field, commonly an electric or magnetic one \cite{GrosjeanLagubeauDarras2015, GrosjeanHubertLagubeau2016, DreyfusBaudryRoper2005}. Therefore, we first consider here a toy model swimmer that is subject to an \textit{external} force field which shall act in one direction only (without restriction of generality the $x$-direction) for the sake of simplicity. For the swimmer to be self-propelled, we demand that all forces acting on the three beads sum up to zero and also have vanishing total torque. 
We determine the remaining degree of freedom by prescribing that the sum of the squares of all forces is equal to a constant, $(2A)^2$, such that the forces explicitly are given by 
\begin{align}
\vec{E}_1 (t, z_1, z_2, z_3) = c \vec{e}_x (z_3 - z_2) \sin (\omega t), \nonumber \\
\vec{E}_2 (t, z_1, z_2, z_3) = c \vec{e}_x (z_1 - z_3) \sin (\omega t), \nonumber \\
\vec{E}_3 (t, z_1, z_2, z_3) = c \vec{e}_x (z_2 - z_1) \sin (\omega t),
\label{eq:DrivingTriangular}
\end{align}
with 
\begin{equation}
c = \frac{(2A)^2}{\sqrt{(z_3 - z_2)^2 + (z_1 - z_3)^2 + (z_2 - z_1)^2)}}. 
\end{equation}
Notably, the forces depend explicitly on the configuration of the swimmer, in order to satisfy the force-free and torque-free condition throughout the whole swimming stroke. 

\begin{figure}[h!]
\centering
\includegraphics{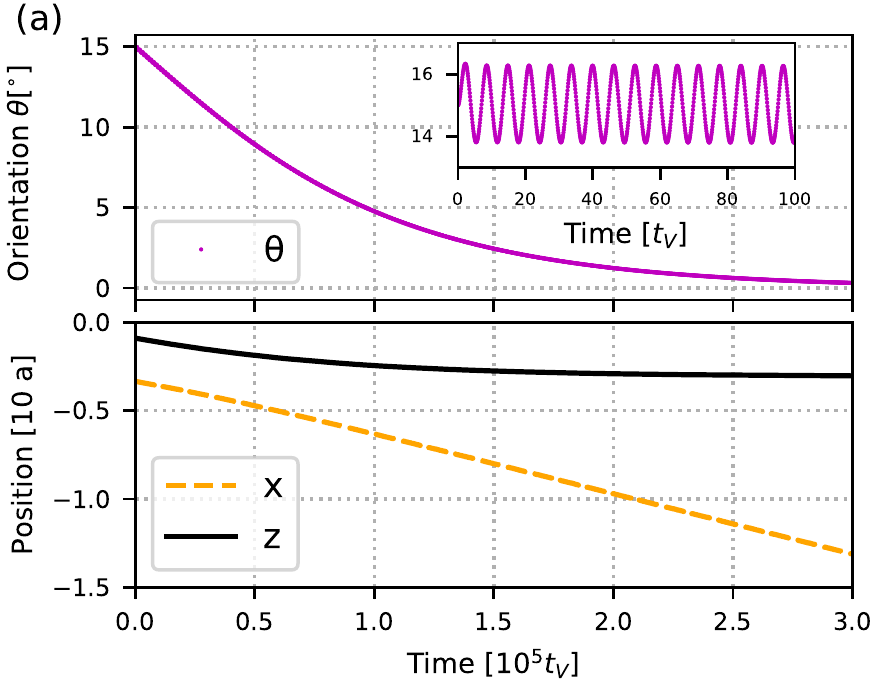}
\includegraphics[scale=1]{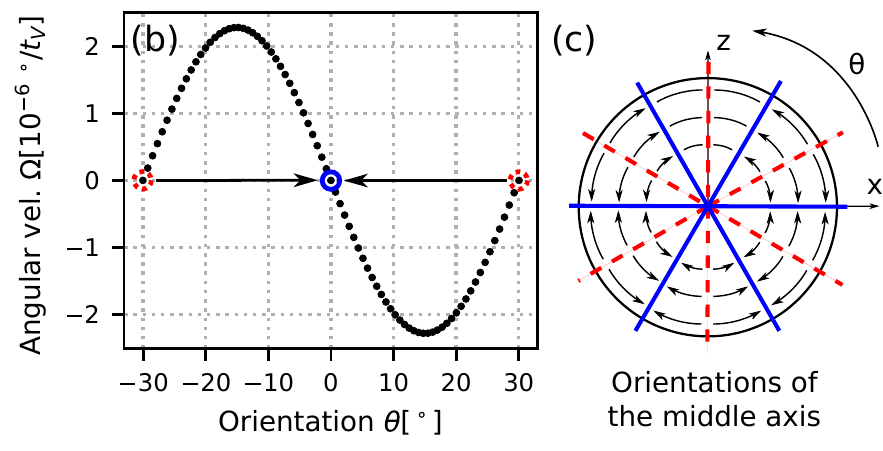}
\caption{Rotational relaxation of the equilateral triangular swimmer in external driving. (a) The time-dependent average orientation (short-time behavior in the inset) and position of the swimmer.  
(b) Angular velocity in dependence of the orientation of the triangular swimmer. 
(c) Sketch of the stable (blue solid) and unstable (red dashed) steady states, where the lines drawn represent the swimmer's middle axis (see figure \ref{fig:sketchTriangularSwimmer}).}
\label{fig:trajectoryRelaxation}
\end{figure}

In numerical studies, we observe that the swimmer typically undergoes a transient phase during which it both rotates and translates. It finally reaches a steady state in which the motion is purely translational (figure~\ref{fig:trajectoryRelaxation} a). 
A closer numerical investigation shows that the angular velocity of the swimmer depends in sinusoidal fashion on the instantaneous orientation of the swimmer (figure~\ref{fig:trajectoryRelaxation} b). The rotational dynamics of the externally driven triangular swimmer hence is equivalent to the one of an overdamped pendulum. Perturbation theory and numerics consistently show that the angular velocity $\Omega$ at fixed angle scales as 
\begin{equation}
\Omega \sim \epsilon^4 \, \nu^4
\end{equation}
and attains its maximum close to the inverse viscous time, similarly to the translational velocity. The time scale of the rotational relaxation hence can be estimated as $T_{relax} \sim \Omega^{-1} \sim \epsilon^{-4} \, \nu^{-4}$. In the parameter range for which the perturbation approach applies, this time scale is several orders of magnitude larger than $t_V$. 

We find stable steady states of the swimmer at $\theta = p \cdot 60^\circ$ and unstable steady states at $\theta = p \cdot 60^\circ - 30 ^\circ,\ p = 1, ..., 6$ (see figure~\ref{fig:trajectoryRelaxation} c). Obviously, the swimmer in external driving is invariant under a 3-fold rotation. Furthermore, a $180^\circ$ rotation inverts the swimming direction but does not affect the stroke in internal coordinates, showing that the stability of states is invariant under a 6-fold rotation. Hence, all stable steady states can be considered equivalent and likewise all unstable states. 
The swimmer is found to always rotate towards the stable steady state closest to the initial orientation as shown in figure~\ref{fig:trajectoryRelaxation} b. 
The existence of stable and unstable steady states results from prescribing the driving forces with respect to the laboratory frame compared to their prescription in the swimmer's frame of reference \cite{RizviFarutinMisbah2018}. In the latter protocol, the forces are held constant in the internal coordinates throughout the rotation of the swimmer and hence the swimmer undergoes constant translation and rotation. 

\begin{figure}
\centering
	\begin{subfigure}[t]{0.49\columnwidth}
	\centering
		\includegraphics[scale=1]{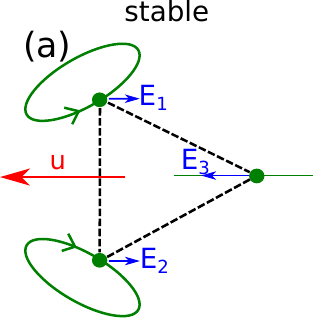}
	\end{subfigure}
	\begin{subfigure}[t]{0.49\columnwidth}
	\centering
		\includegraphics[scale=1]{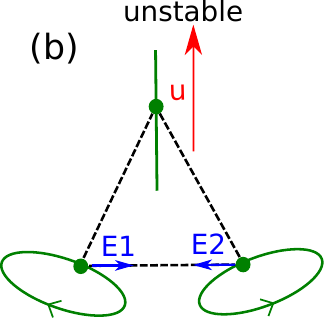}
	\end{subfigure}
	
	\begin{subfigure}[t]{0.49\columnwidth}
	\centering
		\includegraphics[scale=1]{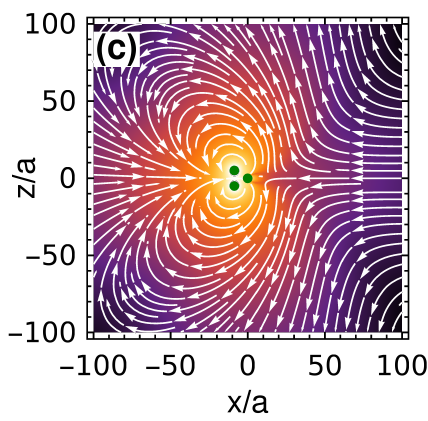}
	\end{subfigure}
	\begin{subfigure}[t]{0.49\columnwidth}
	\centering
		\includegraphics[scale=1]{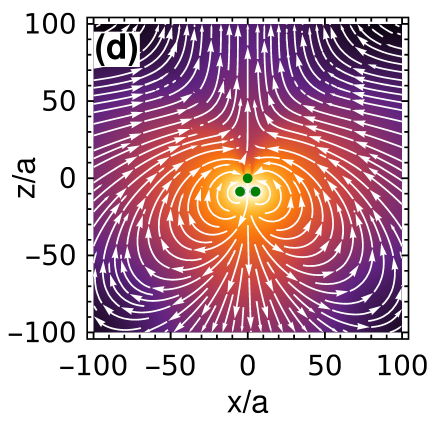}
	\end{subfigure}
	
	\begin{subfigure}[t]{0.49\columnwidth}
	\centering
		\includegraphics[scale=1]{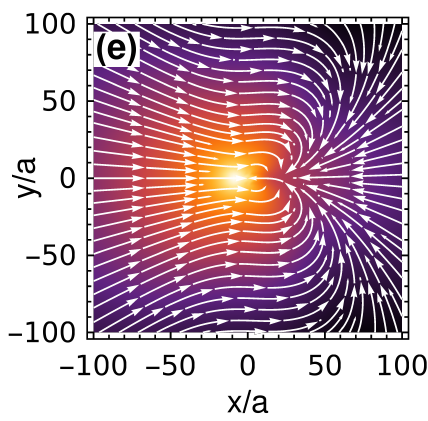}
	\end{subfigure}
	\begin{subfigure}[t]{0.49\columnwidth}
	\centering
		\includegraphics[scale=1]{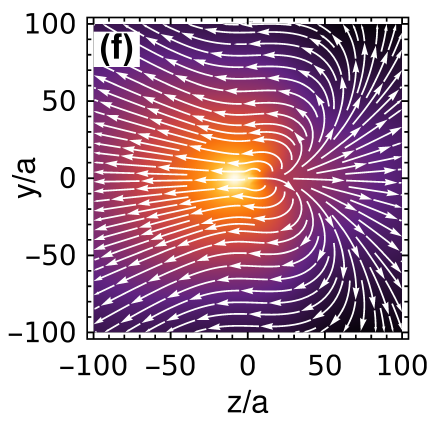}
	\end{subfigure}
\caption{Bead trajectories and flow fields of the triangular swimmer in stable and unstable states. Sketch of the bead trajectories and swimming velocity in the (a) stable and (b) unstable steady state. 
Flow fields in the swimmer plane for the (c) stable and (d) unstable state and flow fields in the orthogonal plane containing the swimmer's symmetry axis for the (e) stable and (f) unstable state. The color scale indicates the magnitude of the flow field. 
}
\label{fig:triangularFlow}
\end{figure}

In the steady states of the externally driven swimmer, we find for the swimming velocity in the Oseen approximation the following expression
\begin{equation}
u_x^{\mathrm{O, tri}} = -\frac{a}{t_V} \frac{12 \sqrt{3} \epsilon^2 \Gamma ^2 \nu ^2 \left(2835 \nu ^3-8640 \nu ^2+5568 \nu -1024\right)}{(9 \nu -8)
   \left(64 \Gamma ^2+9 (8-15 \nu )^2\right) \left(256 \Gamma ^2+9 (8-9 \nu )^2\right)}.
\label{eq:triangularVelocity}
\end{equation}
Performing an expansion and truncation of~\eqref{eq:triangularVelocity} to leading power of $\nu$ shows that this result is in agreement with the result presented in \cite{RizviFarutinMisbah2018}. 
We recover the typical $u \sim \epsilon^2$ dependence and $u \sim \nu^2$ in the limit $\nu \to 0$, which seem characteristic for bead-spring swimmers. We can understand the latter dependence from the fact that swimming emerges from the interplay between the hydrodynamic interaction of parts of the swimmer and variations in their distance, suggesting a swimming speed scaling with the gradient of the hydrodynamic interaction and hence with $\nu^2$. 
In the range of $L > 3a$, we find that expression~\eqref{eq:triangularVelocity} is negative meaning that the swimmer swims towards the base (with respect to the symmetry) of the triangle (see figure~\ref{fig:triangularFlow} a). For sufficiently small amplitudes of the driving, we observe that the beads 1 and 2 move on ellipsoidal trajectories that are tilted towards the middle axis of the swimmer. 
In the orientation $\theta = \pi/2$, exemplaric for the unstable steady states, the swimmer swims at the same speed as in a stable state~\eqref{eq:triangularVelocity}. We point out that this property is sensible to the way the normalization of the forces is done (see ~\eqref{eq:DrivingTriangular}), i. e. it holds only if the sum of the squares is fixed. In contrast to the stable states, the swimming direction is here pointing towards the tip of the triangle and also the trajectories of the beads 1 and 2 are in this case tilted towards the base of the triangle (figure~\ref{fig:triangularFlow} b).

Analysis of the analytically computed flow fields shows that in contrast to previous works, which reported that triangular swimmers produce in average neutral flow fields at second order in the driving force \cite{RizviFarutinMisbah2018, RizviFarutinMisbah2018b}, we here find a non-vanishing average dipolar flow field at order $\epsilon^2$  in both states (see figure~\ref{fig:triangularFlow}  c, d). Going from stable to the unstable state comes with an exact inversion of the flow field transforming the puller-like swimmer in the stable state into a pusher-like swimmer in the unstable one.

\subsection{Internal driving and flow field}
\begin{figure}
\centering
\includegraphics[scale=1]{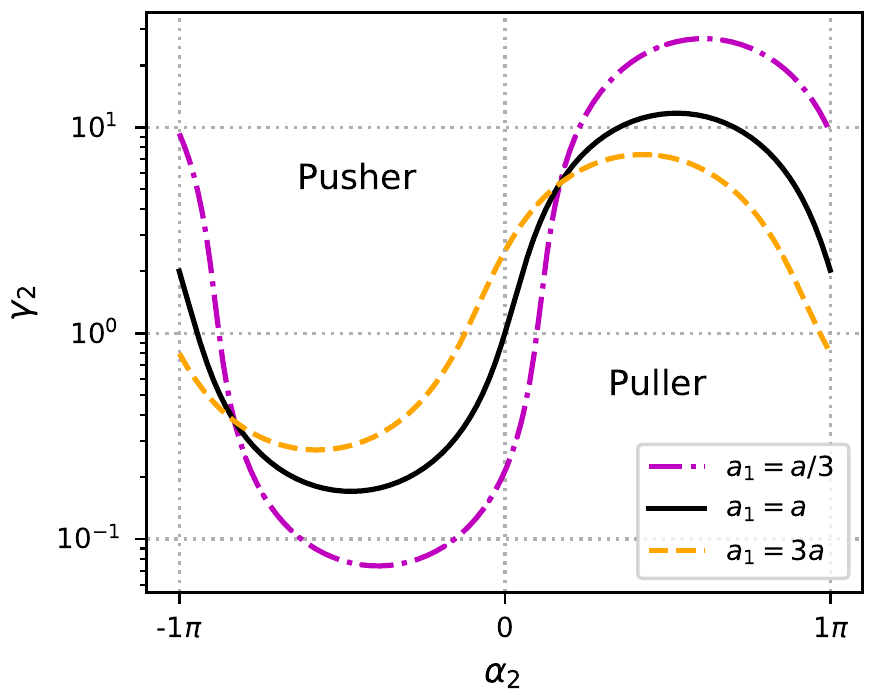}
\caption{
behavior of the equilateral triangular swimmer with variable size of bead 1 in the internal driving in dependence of $\alpha_2$ and $\gamma_2$. Parameters: $\Gamma = 1, \nu = 1/10$.}
\label{fig:PusherPullerTriInternal}
\end{figure}

In order to provide deeper insights into the swimmer's pusher or puller character, we repeat the calculation of the flow field for a purely translational driving protocol presented similarly in \cite{RizviFarutinMisbah2018}. We assume forces $F_{ij}$ between each pair of beads $i, j$ that are defined as
\begin{equation}
F_{12} (t) = F_{13} (t) = A \sin(\omega t), \ F_{23} (t) = \gamma_2 A \sin(\omega t + \alpha_2). 
\end{equation}
The application of the perturbative approach presented here and the analysis of the resulting flow field shows that the fields at order $\epsilon^2$ (in dependence of the rescaled position $\vec{x}'$) can be approximated to leading order as a superposition of two force dipoles
\begin{equation}
\vec{u}^{\mathrm{flow}} (\vec{x}') = f(\alpha_2, \gamma_2) \left( \vec{G}'_D (\vec{x}', \hat{e}_x, \hat{e}_x) - \vec{G}'_D (\vec{x}', \hat{e}_z, \hat{e}_z) \right),
\label{eq:flowTriangularSwimmer}
\end{equation}
with 
\begin{equation}
\vec{G}'_D (\vec{x}', \vec{d}, \vec{e}) := (\vec{d} \circ \nabla' \hat{T}'(\vec{x}')) \circ \vec{e}.
\end{equation}
$\vec{e}$ denotes the direction in which the forces act, $\vec{d}$ the separation vector of the two forces of the dipole and $\nabla'$ the gradient with respect to rescaled coordinates. In the direction orthogonal to the swimmer plane, the two force dipoles cancel up to order $1/y^4$, such that in this direction only the quadrupolar flow field, scaling as $1/y^3$, is reminiscent. 
Investigation of the magnitude of the dipolar flow field shows that a swimmer can be tuned from pusher to puller by changing $\alpha_2$ and $\gamma_2$ (figure~\ref{fig:PusherPullerTriInternal}, black curve). 

Comparison of the strength of the force dipole $f(\alpha_2, \gamma_2)$ produced by the swimmer and the swimming velocity $u_{swim}$ shows that their ratio is a function of $\nu$ only:
\begin{equation}
\frac{f(\alpha_2, \gamma_2)}{u_{swim}} =  -\frac{4 \sqrt{3} (81 \nu -40)}{\nu  \left(135 \nu ^2-360 \nu +128\right)}.
\label{eq:ratioFlowSwimmingVelTriangular}
\end{equation}
As this ratio is positive for physical values of $\nu$, a swimming motion towards the tip of the triangle is always associated to a pusher flow field whereas motion towards the base is associated to a puller character. 
For $\gamma_2 = 1/2$ and $\alpha_2 = \pi$, we recover the driving of the stable steady state, showing that the swimmer behaves as a pusher. The unstable steady state is recovered for $\gamma_2 \to \infty$ making the corresponding swimmer a pusher. Plots of the flow fields in the swimmer's plane and in the orthogonal plane through the symmetry axis of the swimmer (figure~\ref{fig:triangularFlow} c - f) illustrate the dipolar character of the flow in the plane of the swimmer and its fast decay orthogonal to the swimmer's plane. 

The complexity of the flow field at order $\epsilon^2$ increases when the radius of bead 1 is chosen differently, in particular the form~\eqref{eq:flowTriangularSwimmer} does not hold and  the fast decay of the magnitude in $y$-direction is lost. Also the curve determining the pusher/puller behavior in dependence of $\alpha_2$ and $\gamma_2$ is sensitive to changes in the parameter $a_1$ (figure~\ref{fig:PusherPullerTriInternal}). Still, the labels of the pusher and puller area are positioned such that they do not only hold for $a_1 = a$ (solid line) but also for $a_1 = a/3$ (dot-dashed line) and $a_1 = 3$ (dashed line).  

\section{Discussion and Conclusions}
We presented a general perturbative approach to calculate the trajectories, internal dynamics and the flow fields of swimmers consisting of beads interacting by harmonic potentials organized in an arbitrary geometry and subject to a force protocol of choice. We first applied our method to the linear swimmer as a benchmark. Comparison to previous results shows that the qualitative behavior of the swimmer, as presented in earlier works \cite{PandeSmith2015}, is recovered. However strong quantitative differences are found in the case of asymmetric swimmers. We showed that for both, the symmetric and asymmetric linear swimmer, our approach yields a swimming velocity that is correct up to order $\nu^3$, whereas previous approaches are only correct up to order $\nu^2$. This is due to the different ways of extracting the swimming velocity from the constant contribution to the source term. The current formulation maintains consistency and provides corrections that were to the best of our knowledge unaccounted for in previous models. The consequences of these terms are negligible in the case of the symmetric swimmer for which the swimming modes are orthogonal. However, since the deviations from orthogonality increase with enhancing the asymmetry of the linear design, important quantitative differences can be observed in the swimming velocity.  

We further investigated the dynamics of an externally driven triangular swimmer with equal radii and discovered the existence of stable and unstable rotational steady states. We showed that the swimmer propagates at the same speed in both states, but in opposite directions with respect to the symmetry axis of the swimmer. In contrast to previous results \cite{RizviFarutinMisbah2018}, the average flow field produced by the swimmer was shown to be puller- or pusher-like, depending on the forces driving the swimmer. Actually, the character of the dipolar flow field can be directly associated with the swimming direction, which is, unlike in the stroke-based models, a consequence of the driving. Interestingly, the average flow field of the triangular swimmer with equal radii is strong near the swimmer plane, but shows a quick decay in the direction orthogonal to the swimmer plane, in contrast to the linear swimmer which produces a rotationally symmetric flow field. This may prove to be important when considering assemblies of swimmers. 

Apart from investigating planar swimmers composed of more than three beads or three-dimensional bead spring swimmers, the framework presented here is also applicable to an ensemble of interacting microswimmers. With small reformulations, our approach allows for the calculation of relative and absolute translational and angular velocities of a swimmer in the presence of an arbitrary configuration of other ones. The model is, therefore, well suited to provide a comprehensive study of the mechanisms underlying the interaction of microswimmers. While this issue has been addresed in part for dumbbell-shaped swimmers \cite{LaugaBartolo2008, AlexanderYeomans2008}, the stroke-based linear swimmer \cite{PooleyAlexanderYeomans2007, AlexanderPooleyYeomans2008} and recently, for a force-based linear swimmer \cite{KurodaYasudaKomura2019}, a full discussion of the interaction between bead-based microswimmers is still pending. The tools presented herein can be used as the foundation of such analysis, which is a task that we intend to address in future work.

\ack{Acknowledgements}
The authors acknowledge founding by the European Research Council through the grant MembranesAct ERC Stg 2013-337283 and by the DFG Priority Programme “Microswimmers – From Single Particle Motion to Collective behavior” (SPP 1726). 
We also thank Alexander Sukhov, Galien Grosjean, Oleg Trosman and Jayant Pande for valuable discussions. 

\appendix
\section{Velocity of the linear swimmer in the Rotne-Prager approximation}
The result for the swimming velocity of the linear swimmer in the Rotne-Prager approximation is given by 
\begin{equation}
u_Z^{\mathrm{RP, lin}} = \frac{a}{t_V} \epsilon^2 6 \Gamma  \nu ^2 C_0 \left( C_S \sin(\alpha) + C_C \cos(\alpha) + C_K \right), 
\label{eq:swimmingVelLinRP}
\end{equation}
with 
\begingroup\makeatletter\def\f@size{10}\check@mathfonts
\def\maketag@@@#1{\hbox{\m@th\large\normalfont#1}}%
\begin{gather}
C_0 = \frac{\left(930 \nu ^8-2847 \nu ^6+1230 \nu ^5+2736 \nu ^4-2280 \nu ^3-260 \nu^2+792 \nu -224\right)}{\left(31 \nu ^3-42 \nu +24\right) \left(64 \Gamma ^2+\left(\nu ^3-6 \nu +8\right)^2\right) \left(64 \Gamma ^2+\left(31 \nu ^3-42 \nu+24\right)^2\right)}, \nonumber \\
C_S =\left(64 \Gamma ^2+31 \nu ^6-228 \nu ^4+272 \nu^3+252 \nu ^2-480 \nu +192\right), \\
C_C = 16 \Gamma  \left(15 \nu ^3-18 \nu + 8\right), \ C_K = 8 \Gamma  \left(15 \nu ^3-18 \nu +8\right). \nonumber
\end{gather} \endgroup

\section*{References}
\bibliography{Ziegler201810} 
\bibliographystyle{iopart-num} 
\end{document}